# Interfacial-Thermo-Fluid-Adhesion Dynamics of Evaporating Capillary Bridges between Curved Surfaces

**Arnov Paul**[1, a, *], **Subhadeep Mondal**[2], and **Purbarun Dhar**[2, b, *]

[1]Applied Mechanics Department, Motilal Nehru National Institute of Technology Allahabad, Uttar Pradesh- 211004, India

[2] Hydrodynamics and Thermal Multiphysics Lab (HTML), Department of Mechanical Engineering, Indian Institute of Technology Kharagpur, West Bengal – 721302, India

**Corresponding author*:

[a] E-mail: arnov0381@gmail.com ; [b] E-mail: purbarun@mech.iitkgp.ac.in

## Abstract

We probe the evaporation mechanism, and the associated adhesion dynamics of liquid capillary bridges connecting two curved, solid substrates. The coupled thermo-fluid-species transport and the transient evolution of capillary adhesion during evaporation are systematically examined. An accurate, fully coupled transient numerical framework is developed, wherein the equilibrium capillary profiles are first determined from level set method. Next, the evaporation is simulated via Arbitrary Lagrangian-Eulerian (ALE) framework to accurately track the moving liquid-vapor interface. The combined influence of substrate curvature, surface wettability, and solid's thermal conductivity on evaporation and capillary adhesion character is comprehensively analysed. The simulation methodology is robustly validated against published literature for capillary profiles, evaporation rates, and capillary forces, demonstrating good agreement. Our results reveal that the evaporation characteristics of both hydrophilic and superhydrophobic (SH) liquid bridges are strongly governed by substrate's curvature and thermal conductivity, and increasing values pose favourable condition for augmented interfacial mass transfer rate. The innately non-uniform vapour flux generates spatially varying evaporative cooling, producing surface-tension gradients that drive internal thermo-capillary circulation. A non-dimensional scaling analysis shows that Marangoni flow dominates buoyancy-induced flow throughout. Also, increasing substrate curvature decreases the overall capillary force, owing to the reduced curvatures of the liquid bridge, while the temporal evolution of the adhesion force is strongly influenced by both substrate curvature and wettability. Our findings offer valuable insights towards design guidelines for evaporation-driven particle assembly, microgravity experiments, precise atomic force microscopy, bio-sensors design, additive manufacturing, and niche micro-fabrication processes, where the dynamic evolution of liquid-bridge geometry and adhesion critically determines particle packing, structural integrity, and the consequent microstructure.

## 1. Introduction

Capillary bridges (or liquid bridges) often play fundamental roles in thermo-fluidic, and species transport phenomena between adjacent solid substrates. These structures are formed when a small liquid-mass is partially stretched or squeezed between two solid substrates very close to one another, and are relevant to several natural and industrial processes. For instance, the transient changes in mechanical properties of unsaturated soil, wet granular matter, and porous media greatly depend on their liquid retention capability, which in turn is governed by the evaporation behavior of liquid in the form of capillary bridges [1-3]. The hydrodynamics and deformation in seepage flow through unsaturated fractures (in geophysical flows, and oil recovery applications) greatly depend on formation of these bridges [4]. The formation, stability, and evaporation of liquid bridges are relevant to several scientific application such as printing technology [5], trickle bed reactors [6], atomic force microscopy [7], microgravity experiments [8], dispensing of glue in packaging, leather, rubber and vinyl industries [9], heat switch system in electronic cooling [10], and in micro-plating applications [11]. Also, evaporating capillary bridges play a pivotal role in particle assembly and self-assembly by generating capillary forces that drive particle rearrangement and organization into ordered micro- and nanostructures, thereby enabling fabrication of functional materials and coatings [12]. Hence, an in-depth understanding of the interfacial profile evolution and thermo-fluid-species transport behavior of capillary bridges during evaporation under varied circumstances is essential for the rational design and optimization of the aforementioned processes.

Under mechanical equilibrium, the capillary profile is governed by Young-Laplace equation: $\gamma/R_0 = \Delta p + G$; where $\gamma$, $R_0$, $\Delta P$ and $G$ represent liquid-gas interfacial surface tension, overall radius of curvature, Laplace pressure, and gravitational effects. Here, the geometrical parameter $R_0$ takes into account both gorge and external radius of curvature; and thereby is strongly dictated by the wetting properties of solid surface and solid-liquid bridge geometry [13]. For instance, between superhydrophobic (SH) surfaces, both external and gorge profile becomes convex, thereby resulting in a nodoid-shaped (a hyperbolic roulette) bridge; whereas between hydrophilic surfaces a concave gorge and concave external profile lead to an unduloidal (an elliptic roulette) bridge profile [14]. Both these profiles, under respective positive and negative internal pressure conditions, were experimentally realized by Mielniczuk et al [15] under negligible gravity conditions. In presence of gravitational effects, the conventional analytical models cannot be employed to solve Young-Laplace equation due to redundant boundary conditions, and in such cases numerical shooting techniques are needed [16]. Also, the capillary profile also depends on the surface tension at the liquid-gas interface, which in turn is governed by the thermal state of the system. For most liquids, the surface tension decreases with increasing temperature. Consequently, variations in the thermal state (such as arising from external heating or evaporative cooling) can significantly alter the equilibrium capillary profile.

The temporal evolution of capillary profile significantly affects the evaporative transport due to the changing interfacial curvature; as local variation in the curvature of the liquid-vapor interface governs the local vapor concentration gradient in the gaseous domain [17]. The solid-liquid-contact geometry, thermo-physical properties, thermal effects, and ambient conditions [18-21] also govern the capillary evolution. For instance, a larger bridge height results in an enhanced evaporative mass loss rate due to reduced confinement effects [20]. Also, a higher substrate temperature or lower ambient relative humidity substantially enhances the evaporation rate, as reported in recent studies [18, 19]. These parameters also markedly influence the convective transport phenomena within the liquid bridge. The non-uniform evaporative mass flux [22] induces uneven evaporative cooling, thereby generating internal flow through thermal Marangoni convection or buoyancy-driven advection, similar to non-isothermal capillary bridges [23]. Furthermore, confinement effects may significantly modify the internal advection dynamics through changes in the evaporation rate, similar to sessile droplets [24]. Also, in the presence of external confinement, vortex-induced modifications of the gaseous flow field may further alter the evaporation rate [25]. Therefore, the advection characteristics in both the liquid and gaseous domains play a crucial role in determining the overall stability and evaporation behavior of capillary bridges.

Moreover, the capillary force in a liquid bridge arises from the combined contributions of surface tension and Laplace pressure and is therefore strongly governed by the bridge geometry and wetting characteristics [26, 27]. For a given bridge volume, a smaller separation distance increases the interfacial curvature, thereby enhancing both the Laplace pressure and surface tension contributions acting between the substrates [28]. Also, the wettability of the confining surfaces strongly influences the contact angle and contact radius, resulting in predominantly attractive (suction) and repulsive capillary forces for hydrophilic and superhydrophobic (SH) surfaces, respectively [17]. Consequently, the bridge geometry and wetting state play a pivotal role in determining the magnitude and direction of the capillary force. During evaporation, continuous interfacial mass loss alters the transient capillary profile and its characteristic geometrical features, such as the gorge and external radii of curvature. These geometric features are further affected by the contact line dynamics [29] which in turn is dictated by the wetting state and curvature of the solid substrates similar to sessile droplets [30]. As a result, during evaporation, the capillary force initially increases to a maximum until the onset of pressure instability, beyond which it progressively decreases and eventually becomes repulsive (positive gauge pressure) at bridge rupture [28, 29]. Such transient variations in capillary force govern particle rearrangement and self-assembly, enabling the fabrication of ordered micro- and nanostructures for advanced functional materials and coatings.

The foregoing literature highlights the influential role of solid-liquid geometry in governing the capillary profile and evaporation behavior of liquid bridges under varied conditions. It also demonstrates the significant effects of surface wettability and the thermo-physical properties of the solid and liquid phases on the internal thermo-fluidic transport and capillary force dynamics. However, in practical applications, solid substrates are seldom perfectly planar, and curved geometries are more commonly encountered. Such substrate curvature, together with the wetting state, can substantially modify the equilibrium capillary profile by altering the minimum-energy configuration of the whole system. Furthermore, in microgravity experiments involving liquid bridges, the tip material and consequently the thermal conductivity of the confining substrates can vary over a wide range, thereby influencing the evaporation process.

Despite their practical significance, only a few studies [28, 29, 31] have considered the influence of substrate curvature on the evaporation behavior of liquid bridges. Moreover, the coupled effects of evaporative cooling, thermo-fluidic transport, and transient adhesion dynamics have remained largely unexplored. To address these gaps, the present study numerically investigates the coupled thermo-fluid-adhesion dynamics of evaporating liquid bridges. For this purpose, the capillary profile is numerically obtained using level-set method, and the evaporative transport phenomena of liquid bridges are subsequently modeled within an Arbitrary Lagrangian-Eulerian (ALE) framework. The coupled effects of substrate curvature under different wettability and thermal conductivity conditions are explored systematically. The findings provide fundamental insights for optimizing evaporation-driven particle assembly and additive manufacturing processes, where the transient evolution of liquid-bridge geometry and capillary forces governs particle packing, structural integrity, and the resulting microstructure. Furthermore, the mechanistic understanding of evaporation-driven adhesion dynamics developed in the present study provides a scientific basis for improving the quantification of capillary adhesion, tip-sample interaction forces, and nanoscale interfacial phenomena in atomic force microscopy (AFM) across a broad range of operating conditions [32, 33].

## 2. Physical system, mathematical description, and computer simulations

We consider the evaporation phenomenon of a capillary bridge between the curved faces of two adjacent cylindrical substrates in an initially quiescent and isothermal environment. The problem involves coupled mass, momentum, and energy transport between the adjacent solid, liquid and gaseous domains. At initial conditions, the liquid molecules being saturated at the liquid-vapor interface diffuse to the surrounding gaseous domain solely due to concentration difference. During this process, latent heat is extracted from the interface, resulting in evaporative cooling and the development of temperature gradients within both the liquid and gaseous phases. These thermal disturbances further propagate into the adjacent solid domains through heat conduction. This thermal non-uniformity induces bulk motion in both fluid and gaseous domain through thermal Marangoni effects or buoyancy driven advection. Such bulk fluid motion may result in internal heat advection thereby the affecting the saturation concentration of vapor at the liquid-vapor interface. In other words, the evaporation effects i.e., the evaporative cooling and resulting bulk fluid motion directly influences its driving mechanism i.e., the interfacial vapor diffusion. As a result, the mass, momentum and energy transport processes become strongly coupled and inherently nonlinear. These phenomena may further be affected by the system geometry and wetting effects which in turn is influenced by the thermo-physical properties of participating domains.

The physical problem is formulated mathematically in an axisymmetric *r-z* coordinate system, as shown schematically in fig. 1(a). The liquid bridge is formed between the curved faces of two adjacent cylindrical substrates possessing identical curvature, with the contact diameter being equal to the substrate diameter, as illustrated in Fig. 1(a, b). The corresponding geometrical specifications at the interface (such as gorge radius and external radius of curvature) and at the contact line (such as contact angle, radius of curvature at solid substrate) are illustrated in fig. 1(b). This configuration highlights the intrinsic influence of substrate curvature on the internal thermo-fluid dynamics while minimizing its effect on the surrounding vapor concentration field. To investigate the role of curvature, the curvature magnitude of both the upper and lower substrates is varied over a wide range, as summarized in Table 1. For wettability effects, we consider the solid-liquid contact angle $\theta = \pi/4$ and $\theta = 5\pi/6$ representing hydrophilic and SH nature of the surfaces. Furthermore, aluminum (k = 238 W/m.K), alumina (k = 27 W/m.K), glass (k = 1.4 W/m.K) and acrylic plastic (k = 0.18 W/m.K) are selected as substrate materials to elucidate the influence of solid thermal conductivity relative to that of the liquid domain. For all cases, the initial volume ($V_0$) and axial height ($h$) of the water liquid bridges are considered as 2μl and 0.6mm respectively. For the evaporation study, we considered constant contact radius (CCR) mode of evaporation i.e., the contact lines at the both top and bottom substrate remains fixed while the transient contact angles decrease consistently. Under these conditions, the evaporative transport phenomena are numerically simulated until the bridge volume decreases to 25% and 20% of its initial value for $\theta = \pi/4$ and $\theta = 5\pi/6$ respectively. Beyond these limits, the contact line de-pins with complex stick-slip behavior as reported in previous studies [17, 20, 28, 29].

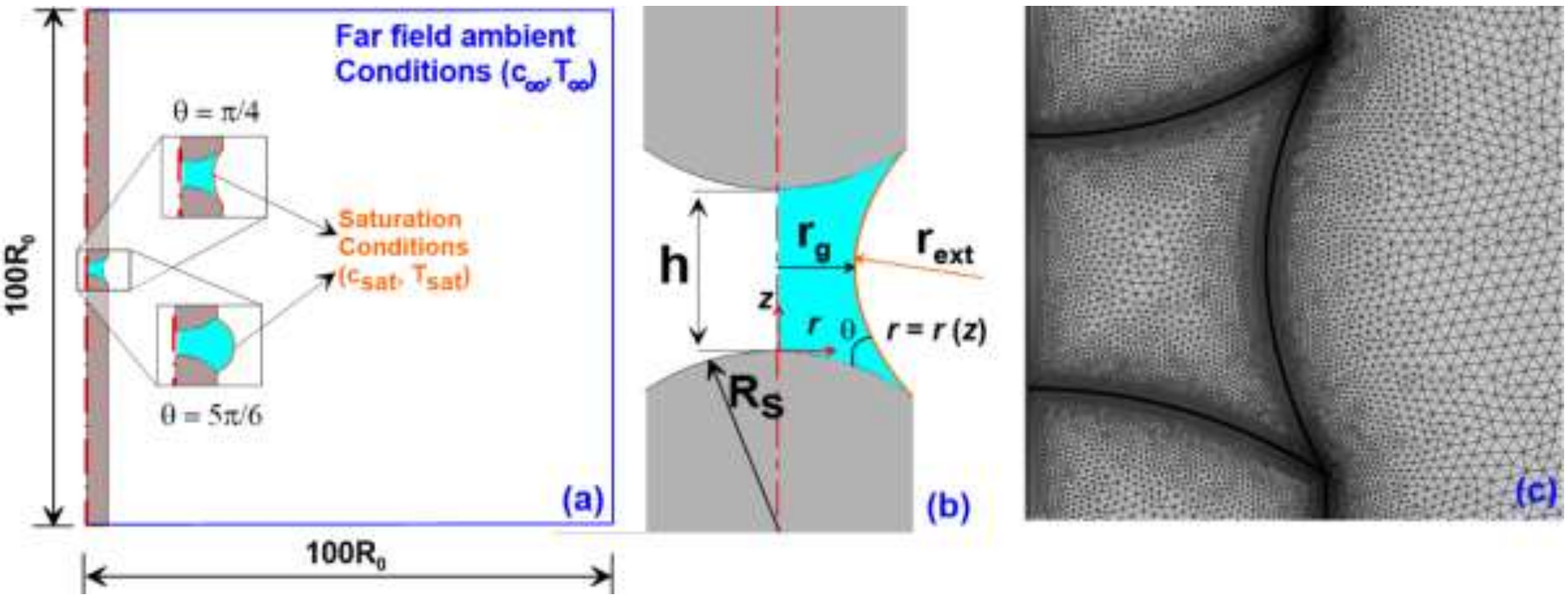

**Fig. 1: (a) schematic of the computational domain for evaporation study; (b) geometry of the capillary bridge between two spherical substrates; (c) typical distribution of mesh structures near liquid-vapor interface.**

**Table 1: Geometrical specification of the substrate curvature considered in present study**

| Nature of curvature | Sl. No. | Curve radius ($R_s$) (mm) | Curvature ($C=R_s^{-1}$) (mm$^{-1}$) |
|---|---|---|---|
| Flat substrate | 0 | ∞ | $C_0 = 0$ |
| Curved | 1 | 10 | $C_1 = 0.1$ |
| | 2 | 3.333 | $C_2 = 0.3$ |
| | 3 | 2 | $C_3 = 0.5$ |

**2.1 Governing Equations:**

**2.1.1 Initial equilibrium profile of the capillary bridges:**

The stable profile of capillary bridges at initial state is obtained using level set (LS) method. In this approach, we define the level set function $\phi$ as the mass fraction of liquid phase and solve the LS equation coupled with the Navier-Stokes equation through the advection velocity given as [34]:

$$\frac{\partial \phi}{\partial t} + \nabla\left(u_{LS}\phi\right) = \xi \nabla \cdot \left(\varepsilon \nabla \phi - \left[\phi\left(1-\phi\right)\frac{\nabla \phi}{\left|\nabla \phi\right|}\right]\right) \tag{1}$$

$$\rho\left(\frac{\partial u_{LS}}{\partial t} + (u_{LS}.\nabla)u_{LS}\right) = \nabla \cdot \left[-pI + \mu\left(\nabla u_{LS} + (\nabla u_{LS})^T\right)\right] + \rho g \tag{2}$$

In these eqns. $u_{LS}$ denotes the velocity in level set method, $\varepsilon$ is the parameter controlling interfacial thickness, $\xi$ is the re-initialization velocity and $g$ is the gravitational acceleration. Typically, $\varepsilon$ depends on the mesh size and $\xi$ is approximated as maximum velocity occurring in the computation domain. To avoid an abrupt discontinuity of the thermo-physical properties at the interface smeared out Heaviside function is used given by $\rho = \rho_l + \left(\rho_v - \rho_l\right)H\left(\varphi\right)$. Here $H\left(\zeta\right)$ is the Heaviside unit step function. These eqns. are numerically solved by imposing the equilibrium contact angle as a wetted wall boundary condition for a fixed liquid volume to obtain the stable capillary profile under varied conditions. The value corresponds to $\phi = 0.5$ always points the position of liquid-vapor interface.

### 2.1.2 Evaporative transport phenomena:

The evaporation behavior of liquid bridges is physically modelled as mass, momentum and energy transfer across the solid, liquid, and gaseous domains. These coupled transport phenomena are mathematically modelled using advection-diffusion eqn., continuity and Navior-Stokes eqn., and energy eqn. given as:

$$\frac{\partial c}{\partial t}+(u_c.\nabla)c=\nabla\cdot\left[D\nabla c\right] \tag{3}$$

$$\nabla.U=0 \tag{4}$$

$$\rho\left(\frac{\partial U}{\partial t}+(u_c.\nabla)U\right)=\nabla\cdot\left[-pI+\mu\left(\nabla U+(\nabla U)^T\right)\right]+G \tag{5}$$

$$\rho C_p\left(\frac{\partial T}{\partial t}+(u_c.\nabla)T\right)=\nabla\cdot\left[k\nabla T\right] \tag{6}$$

Here, $c$, $D$, $U$, $\rho$, $p$, $\mu$, $C_p$, $T$ represent vapor concentration, diffusivity of water vapor, velocity, density, pressure, viscosity, specific heat and temperature respectively. It should be noted that, in the present study, eqns. (3)-(6) are formulated within an arbitrary Lagrangian-Eulerian (ALE) framework (discussed later) to accurately track the moving liquid-vapor interface during the evaporation process. Within this framework, Eq. (3) is solved only in the gaseous domain, whereas Eqs. (4) and (5) are solved in both the liquid and gaseous domains and eqn. (6) is solved in the liquid and solid domains, with the convective transport term neglected in the solid owing to the absence of bulk molecule motion. Also, it is worthwhile to mention that for the convective transport, we employed the convection velocity $u_c$ in ALE framework defined as the difference between material velocity and mesh velocity, i.e., $u_c=u_{fluid}-u_{mesh}$. The transient term ($\left(\partial/\partial t(\ )\right)$ in eqns. (3), (5), and (6) represents temporal change in the mesh coordinate system of ALE framework. The gravitational effects ($G=\left(\rho_r+\Delta\rho\right)g$) in the Navior-Stokes equation are considered using spatial variation of density as a function of temperature for the liquid domain. In case of gaseous domain, the density is a function of both temperature and vapor concentration given as:

$$\Delta\rho_w=\beta_T\rho_{w,r}\left(T-T_r\right) \tag{7}$$

$$\Delta\rho_a=\left(\frac{p_aM_a}{RT}+cM_w\right)-\left(\frac{p_{a,r}M_a}{RT_r}+c_rM_w\right) \tag{8}$$

$$p_a=p_o-cRT \tag{9}$$

Where $\Delta\rho$ , $\beta_T$, $M$, $R$ represents the density change, volumetric expansion coefficient of water, molar mass, and universal gas constant, respectively. The subscripts '*w*', '*a*', and '*r*' denotes water, air and reference state, respectively. Here, the temperature and vapor concentration of the far field are considered as the reference conditions for temperature and vapor concentration. Also, in present study, we have considered the Boussinesq approximation to model the natural convection phenomenon similar to sessile droplet evaporation studies [35].

## 2.2 Initial and boundary conditions:

### *2.2.1 Wetted wall constrains for capillary profile:*

We apply wetted wall constrains i.e., no penetration boundary condition with slip effects at the curved solid surfaces to obtain the initial stable capillary profile using level set method. In physical sense, the contact line slips over the solid geometry to reach a stable state depending upon the specified contact angle at the solid-liquid interface. Mathematically, these boundary constrains can be modelled as [36, 37]:

$$u_{LS} \cdot \vec{n}_{wall} = 0 \tag{10}$$

$$\sigma_{nt} = \sigma_n - \left(\sigma_n \cdot n_{wall}\right) n_{wall} = \gamma\delta\left(\vec{n}_{\text{int}} \cdot \vec{n}_{wall} - \cos(\theta_0)\right)\left(\vec{n}_{\text{int}} \cdot \vec{t}_{wall}\right)\vec{t}_{wall} - \frac{\mu}{\beta} u_{slip} \tag{11}$$

$$u_{slip} = u_{LS} - \left(u_{LS} \cdot n_{wall}\right) n_{wall} \tag{12}$$

$$\vec{n}_{wall}\left(\varepsilon\nabla\phi - \phi(1-\phi)\frac{\nabla\phi}{|\nabla\phi|}\right) = 0 \tag{13}$$

Here, $\sigma_{nt}$, $\sigma_n$, $u_{slip}$, $\beta$ denotes the tangential stress, normal component of viscous stress tensor, slip velocity and slip length respectively. Also, $\vec{n}_{wall}, \vec{n}_{int}$ represents the unit vector normal to wall and interface and $\vec{t}_{wall}$ is the unit vector tangential to the wall. In present study, the slip length is chosen to be equal to the minimum mesh element size. This condition ensures adequate numerical accuracy while maintaining computational efficiency [38]. It is worth mentioning that, as the imposed contact angle differs from the angle between the wall normal and the interface normal, the contact line slips along the solid wall, causing the interface to evolve under the advection field. When the instantaneous contact angle at the solid-liquid interface approaches the imposed value, the system reaches an equilibrium state and the resulting liquid-vapor interface profile is used as the initial condition for the evaporation problem.

### *2.2.2 Constraints for evaporative transport phenomena*

We consider a local temperature-dependent saturation condition at the liquid-vapor interface corresponding to Antoine eqn. [39] to model the species transport phenomenon. At the far field, the vapor concentration ($c_{ff}$) is maintained at a constant value corresponding to 50% relative humidity, while a no-penetration (zero normal flux) condition is imposed on the solid boundaries. Mathematically, these boundary conditions are expressed as:

$$c_{sat}(r) = \frac{p_{sat}}{RT_i} = \exp\left[A - \frac{B}{T_i(r) - C}\right] \times \frac{10^6}{RT_i(r)} \tag{14}$$

$$c_{ff} = 0.5 c_{sat}(T_{amb}) \tag{15}$$

$$\left.\frac{\partial c}{\partial n}\right|_{wall} = 0 \tag{16}$$

Here, the coefficients $A$, $B$ and $C$ are component specific constants and for water the corresponding values are given as: $A$ = 9.487, $B$ = 3893K and $C$ = 42.68K [40]. Also, the suffix '$i$' represents the interfacial position. This spatial concentration variation results in interfacial evaporative flux ($J$) for which we apply mass conservation at the capillary surface given as:

$$J = \vec{n} \cdot \left(-D\nabla c + U \cdot c\right) M_w \tag{17}$$

$$U_a - U_w = J\left(\frac{1}{\rho_w} - \frac{1}{\rho_a}\right) \tag{18}$$

The evaporative mass flux gives rise to evaporative cooling in both the liquid and gaseous domains, as the evaporated liquid molecules remove the latent heat of vaporization from the interface. The resulting temperature gradients extend into the adjacent solid domain via heat conduction, leading to a coupled heat transfer process across all three phases. The corresponding thermal boundary conditions are mathematically described as:

At liquid-vapor interface: $n\left(k_a\left(\nabla T\right)_a - k_w\left(\nabla T\right)_w\right) = -JL_h$ (19)

At solid-liquid interface: $n\left(k_s\left(\nabla T\right)_s - k_w\left(\nabla T\right)_w\right) = 0$ (20)

At solid-vapor interface: $n\left(k_s\left(\nabla T\right)_s - k_a\left(\nabla T\right)_a\right) = 0$ (21)

Where, $k$ is thermal conductivity and $L_h$ is the latent heat of vaporization. Also, the subscript '$s$' denotes solid domain. The temperature gradients arising from evaporative cooling generate buoyancy and surface tension-driven flow in the liquid and gaseous domains. The two flow fields are coupled at the liquid-vapor interface through the continuity of tangential and normal stresses, accounting for Marangoni stresses and capillary pressure. Mathematically, the interfacial stress balance is given by:

$$\left(\vec{n}_{\text{int}} \cdot S_g - \vec{n}_{\text{int}} \cdot S_l\right) \cdot \vec{n}_{\text{int}} = \gamma\kappa \cdot \vec{n}_{\text{int}} \tag{22}$$

$$\left(\vec{n}_{\text{int}} \cdot S_g - \vec{n}_{\text{int}} \cdot S_l\right) \cdot \vec{t}_{\text{int}} = -\gamma_T \nabla_g T \tag{23}$$

Where $\nabla_g = (I - \vec{n} \cdot \vec{n}^T)\nabla$, $\gamma$, $S$, $t_{int}$, $n_{int}$, $\kappa$, $\gamma_T$ are the surface gradient operator, surface tension, stress tensor, unit vector along tangential and normal direction at interface, interfacial curvature and temperature derivative of surface tension respectively. In addition, a constant temperature condition of $T_{amb}$ = 20°C is imposed at the far-field boundaries and at the end faces of the solid cylinder. A no-slip boundary condition is enforced at all solid-liquid and solid-gas interfaces. The initial values of velocity, temperature, and relative humidity are considered as 0 m/s, 20$^o$C, and 50% respectively.

***2.3 Computer simulations***

In the present study, the numerical simulations are carried out in two stages. First, the equilibrium capillary profile of the liquid bridge is obtained using the level set method under different substrate curvatures and wettability conditions. For this purpose, the level set governing equations (eqns. (1) and (2)) are solved numerically with appropriate boundary conditions (eqns. (10)-(13)). Initially, a liquid column of prescribed volume ($V_0$ = 2 μL) with an arbitrary shape is considered between the solid domain, and the equilibrium contact angle is imposed at the solid-liquid interface. The interface thickness is chosen to be equal to the maximum mesh size in the subdomains surrounding the liquid-vapor interface [41], while the reinitialization parameter is approximated using the characteristic velocity of the liquid bridge, given by: $v_{char} = h / t_c$ where $t_c = \sqrt{\rho h^3 / \gamma}$ . Based on these considerations, the capillary profile is iteratively computed until the liquid bridge attains its equilibrium configuration. The resulting equilibrium interface profile obtained from the level set simulations is subsequently employed as the initial geometry for the evaporation simulations. This two-stage approach ensures that the evaporation process is initiated from a physically realistic and mechanically equilibrated capillary bridge under the prescribed curvature and wettability conditions.

Next, for the evaporation phenomenon, the transient governing equations are solved in a fully coupled manner within an arbitrary Lagrangian-Eulerian (ALE) framework [42, 43]. In addition to the conventional material and spatial coordinate systems, the ALE formulation introduces a moving mesh coordinate system, in which the governing equations are formulated. This enables the computational mesh to move independently of both the material and spatial coordinates, thereby facilitating accurate tracking of the deforming liquid-vapor interface while maintaining high mesh quality throughout the computational domain. For the present study, evaporation is simulated under the constant contact radius (CCR) mode. Accordingly, the mesh at the solid-liquid interface is fixed, whereas the mesh along the liquid-vapor interface is allowed to deform in accordance with the interface motion, enabling precise tracking of the moving interface. However, this independent mesh motion may lead to mesh distortion, thereby deteriorating numerical accuracy. To mitigate this issue, an automatic remeshing strategy is employed, whereby the computational domain is regenerated whenever the mesh quality falls below a prescribed threshold of 0.5.

In present study, both stages of the numerical simulations were carried out using the transient, fully coupled finite element-based solver provided in COMSOL Multiphysics. The computational domain was assumed to be a two-dimensional axisymmetric square with side $100R_0$ as illustrated in Fig. 1(a). We discretise the entire simulation domain using linear Lagrange finite elements for all dependent variables. For meshing, unstructured triangular mesh was employed with the most refinement being concentrated locally near the liquid-vapor interface and the contact line (approximately 100 times smaller than the liquid bridge height; see fig. 1(c)). The time stepping for transient study was obtained using a backward differentiation formula (BDF) scheme, with the maximum Courant number being limited to 0.2. A relative convergence tolerance of 0.005 was prescribed for all dependent variables to ensure numerical convergence. Furthermore, streamline diffusion stabilization was employed to suppress numerical oscillations and improve the stability of the finite element solution, particularly in convection-dominated regions. For grid independence test, we vary the grid points from ~2,25,854 to ~3,74,398 and observed that the change in evaporation rate was varued by only 0.8%.

For validation of present numerical schemes, we compare the results obtained from present simulations with that of previous research works under varied conditions. For instance, the numerically obtained stable capillary profile at equilibrium state between solid spheres is compared with that of Wang et al. [44] for varied Bond number conditions. For the mass loss rate, the transient volume of liquid bridges evaporating between glass spheres is compared against the experimental results reported by Boulodenine et al. [31]. We also compare the transient capillary force due to relative volume loss of evaporating capillary bridges with the experimental results of Mielniczuk et al. [29] for varied heights. It can be observed that the results from present numerical models closely mimic various aspects of capillary bridge evaporation phenomenon quite well except for evaporation rate at later stages of evaporation. This discrepancy can be primarily attributed to contact-line depinning during the experiments, which causes the evaporation process to deviate from the assumed constant contact radius (CCR) mode adopted in the simulations. Nevertheless, the overall agreement between the numerical predictions and experimental observations demonstrates that the present modeling framework accurately captures both the stable capillary profile and the coupled evaporative transport phenomena under varied conditions.

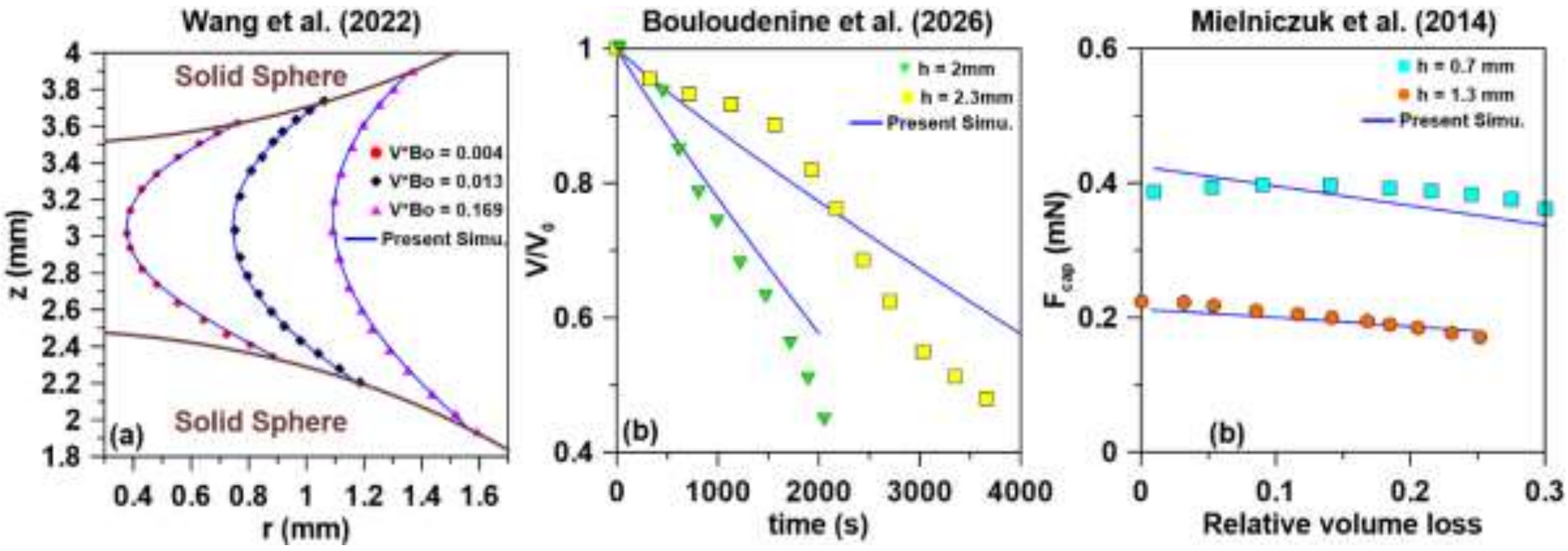

**Fig. 2: Validation of (a) equilibrium capillary profile, (b) transient volume loss, (c) capillary force during evaporation of liquid bridges over curved substrates for varied test conditions.**

## 3. Results and Discussions

### 3.1 Stable capillary profiles at the initial equilibrium state

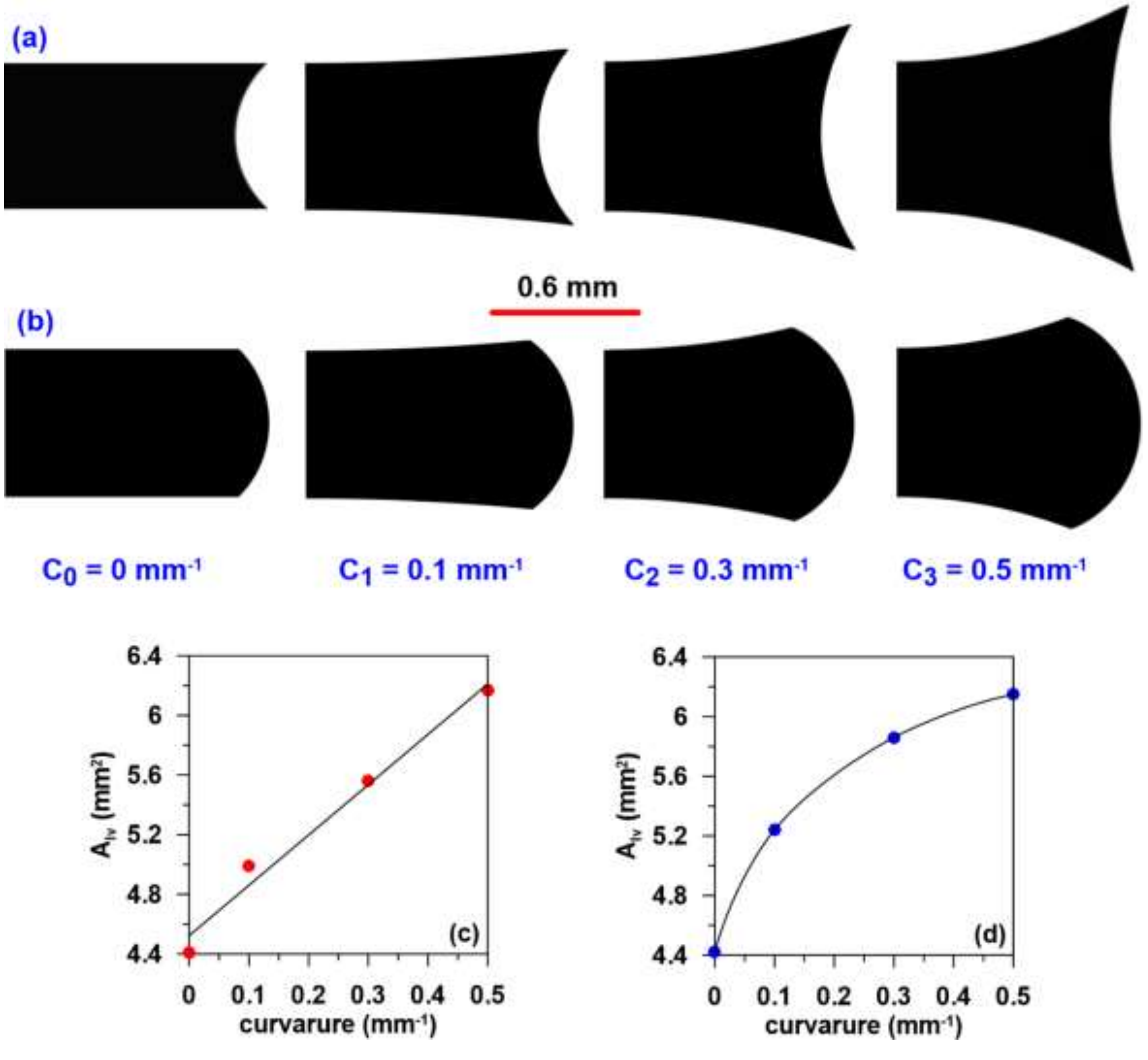

**Fig. 3: Stable capillary profile of liquid bridges at initial equilibrium state for (a) hydrophilic ($\theta = \pi/4$) and (b) SH ($\theta = \pi/4$) surfaces under varied curvature of adjacent substrates [The scale bar in red colour represents 0.6 mm length]; The variation of liquid-vapor interfacial temperature under varied curvature of solid substrates for (c) hydrophilic and (d) SH surfaces.**

The initial stable capillary profiles of liquid bridges at the equilibrium state under varied wettability and curvature conditions are shown in fig. 3 (a, b). The corresponding area of the liquid-vapor interface is shown in fig 3 (c) and 3(d) for respectively hydrophilic and SH surfaces. We notice that for both hydrophilic and SH surfaces the initial equilibrium capillary profile is strongly dictated by the curvature of the substrates between which the bridge is formed. For instance, both in case of hydrophilic and SH surfaces, the external radius of liquid-vapor interfacial curvature is increased with the enhanced curvature of solid substrates;

whereas the gorge radius is reduced. Now, although the influence of substrate curvature on the equilibrium capillary profile appears qualitatively similar for both wettability conditions, its effect on the subsequent transport phenomena may become different. For hydrophilic surfaces, increasing substrate curvature shifts the liquid-vapor interface towards the vertical plane passing through the upper and lower three-phase contact lines. In contrast, for SH surfaces, the effect is quite opposite, with the liquid-vapor interface moving away from the same vertical plane. We also notice that for both hydrophilic and SH surfaces, the enhanced curvature of solid substrate consistently augments the liquid-vapor interfacial area. These curvature-induced modifications to the interfacial geometry substantially influence the evaporative mass flux, thermofluidic transport, and adhesion characteristics during the considered stages of evaporation as discussed in the sub-sequent subsections.

**3.2 Evaporative mass transfer**

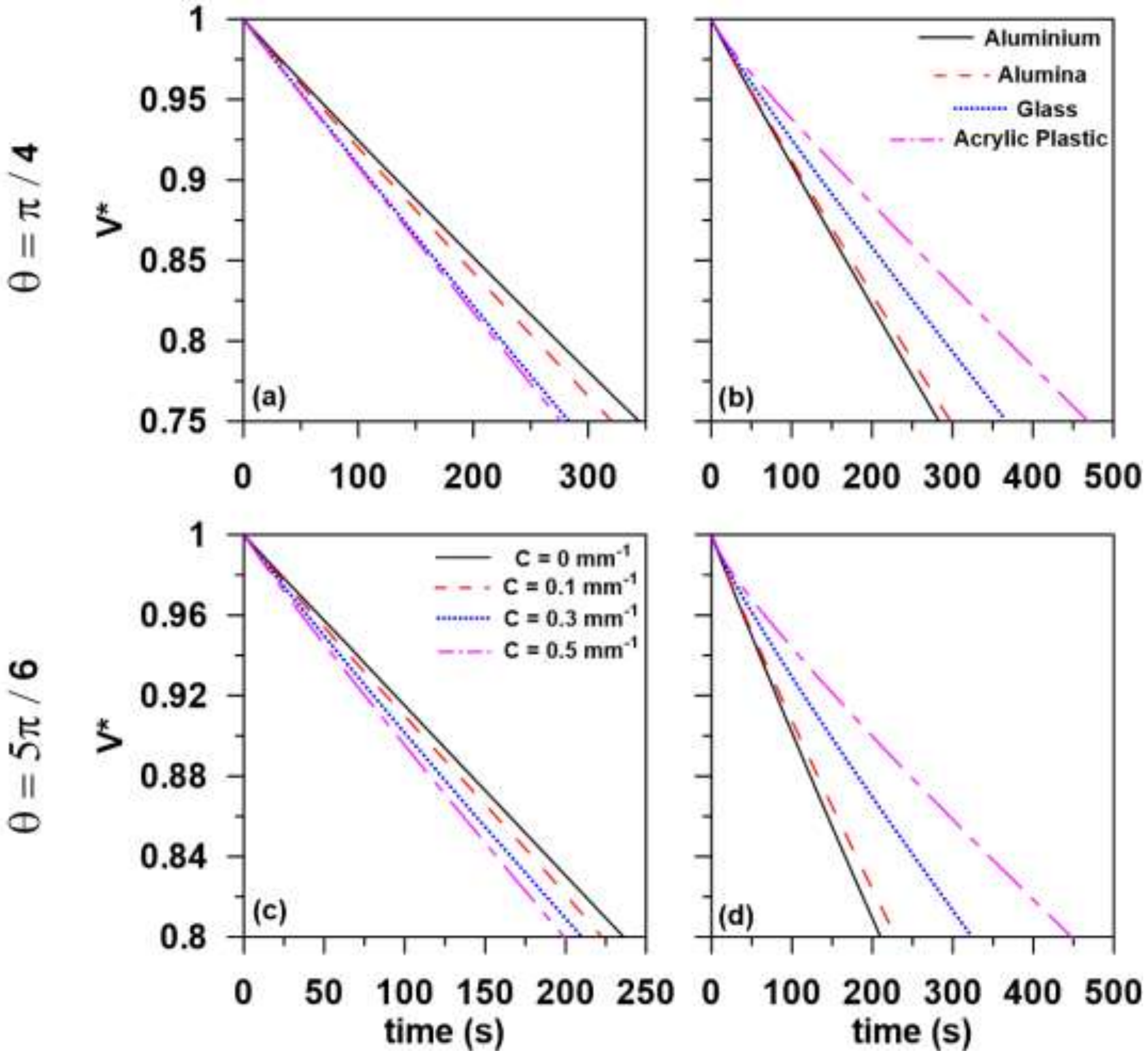

**Fig. 4: Transient evolution of liquid bridge volume during evaporation under varied curvature and thermal conductivity of the solid substrate materials for hydrophilic (a, b) and SH surfaces (c, d) [The legends are same for (a), (c) and (b), (d)].**

The temporal evolution of capillary bridge volume during evaporation under varied curvature and thermal conductivity of solid substrate are illustrated in fig 4. We notice that, for both hydrophilic and SH cases, the slope of the $V^*$ ($=V/V_0$) with time i.e., the evaporation rate is enhanced for curved substrate compared to the flat substrate case and the same is increased in a consistent manner with the augmented curvature of the adjacent substrates. These findings can be attributed to the modified liquid-vapor interfacial geometry brought about the solid substrate curvature. For hydrophilic surfaces, the capillary profile remains concave thereby causing the liquid-vapor interface to remain partially confined between the solid substrates. This condition may result in formation of vapor shroud thereby increasing the local vapour concentration surrounding the liquid-vapor interface. As a result, the vapor concentration gradient is reduced which leads to low evaporation rate for high concave curvature. Now, in case of curved geometry of the solid substrates, the concave nature of the interface is reduced (as seen in fig. 3a) thereby causing a more even spreading of concentration contours around the capillary surface.  This condition augments the local vapor concentration gradient thereby causing higher evaporation rate with enhanced curvature of the solid substrates. For SH surface, the interfacial curvature remains convex for the considered cases. In such cases, the induction of solid substrate curvature causes the liquid bridge to protrude further thereby causing a shorter average diffusion length scale. This condition favours the evaporated molecules to diffuse to far field at a faster rate thereby amplifying the mass loss rate with enhanced curvature. In addition, for both hydrophilic and superhydrophobic (SH) surfaces, the introduction of substrate curvature increases the exposed liquid-vapor interfacial area available for mass transfer to the surrounding ambient for same liquid bridge volume. This enlarged evaporative interface further enhances the overall mass loss rate for curved substrate configurations compared to the flat one.

We also observe that, for a specific geometry (i.e., same solid substrate curvature $c_2 = 0.3\ mm^{-1}$), the evaporation rate consistently decreases with the reduction in thermal conductivity of the solid substrate material for both hydrophilic and SH surfaces. These findings can be attributed to the strong dependency of the saturation concentration on the evaporative cooling induced temperature drop along the liquid-vapor interface. During evaporation, the liquid domain cools down due to the extraction of latent heat by the evaporated liquid molecules. Now, in case of higher thermal conductivity of the adjacent solid substrates, this evaporation induced cooling is partially compensated by the higher incoming heat flux from the solid substrates. In contrast, for lower thermal conductivity cases, the evaporative cooing effects spread across both liquid and solid domains owing to limited heat conduction in solid. This condition results in higher temperature drop i.e., effectively lower temperature along the liquid-vapor interface resulting in a lower interfacial saturation vapor concentration. As a result, the interfacial vapor concentration gradient is reduced, thereby decreasing the evaporation rate compared to substrates with higher thermal conductivity.

Next, we look into the variation of evaporative mass flux along the liquid-vapor interface of capillary bridges under varied conditions considered in present study. Fig. 5 illustrates this mass flux distribution for varied curvature and thermal conductivity of the solid substrate. Here, the evaporative flux is non-dimensionalized as: $J^* = J \Big/ \dfrac{D\left(c_{sat}(T_s) - c_\infty\right)}{R_0}$; where $R_0$ is the overall radius of interfacial curvature for corresponding cases. We notice that, for all considered cases, the evaporative flux tends to diverge near both the contact lines. This finding may be attributed to the reduced probability of vapor re-adsorption due to the presence of the solid boundary. In general, for equilibrium phase transfer phenomenon (such as evaporation), both the phases co-exist at the interface with spontaneous conversion from one to another phase. In other words, during evaporation, the liquid molecules are evaporated and again re-adsorbed at the liquid-vapor interface. Now, near the edge of liquid domain, the probability of re-adsorption is reduced due to the presence of solid boundary. Also, the large diffusion angle between the liquid vapor interface and solid domain provide more scope for the evaporated molecules to diffuse to far field. As a result, the overall diffusion length becomes significantly smaller thereby furnishes steep vapor concentration gradient similar to sessile droplets [45]. This sharp concentration gradient justifies the diverging nature of the evaporative flux near three-phase contact point of the capillary bridges especially for hydrophilic surfaces. In case of SH cases, the interfacial mass flux distribution is relatively more uniform i.e., the diverging nature is partially mitigated due to reduced wedge angle (i.e., shrunk diffusion domain) between the liquid-vapor interface and the solid surface.

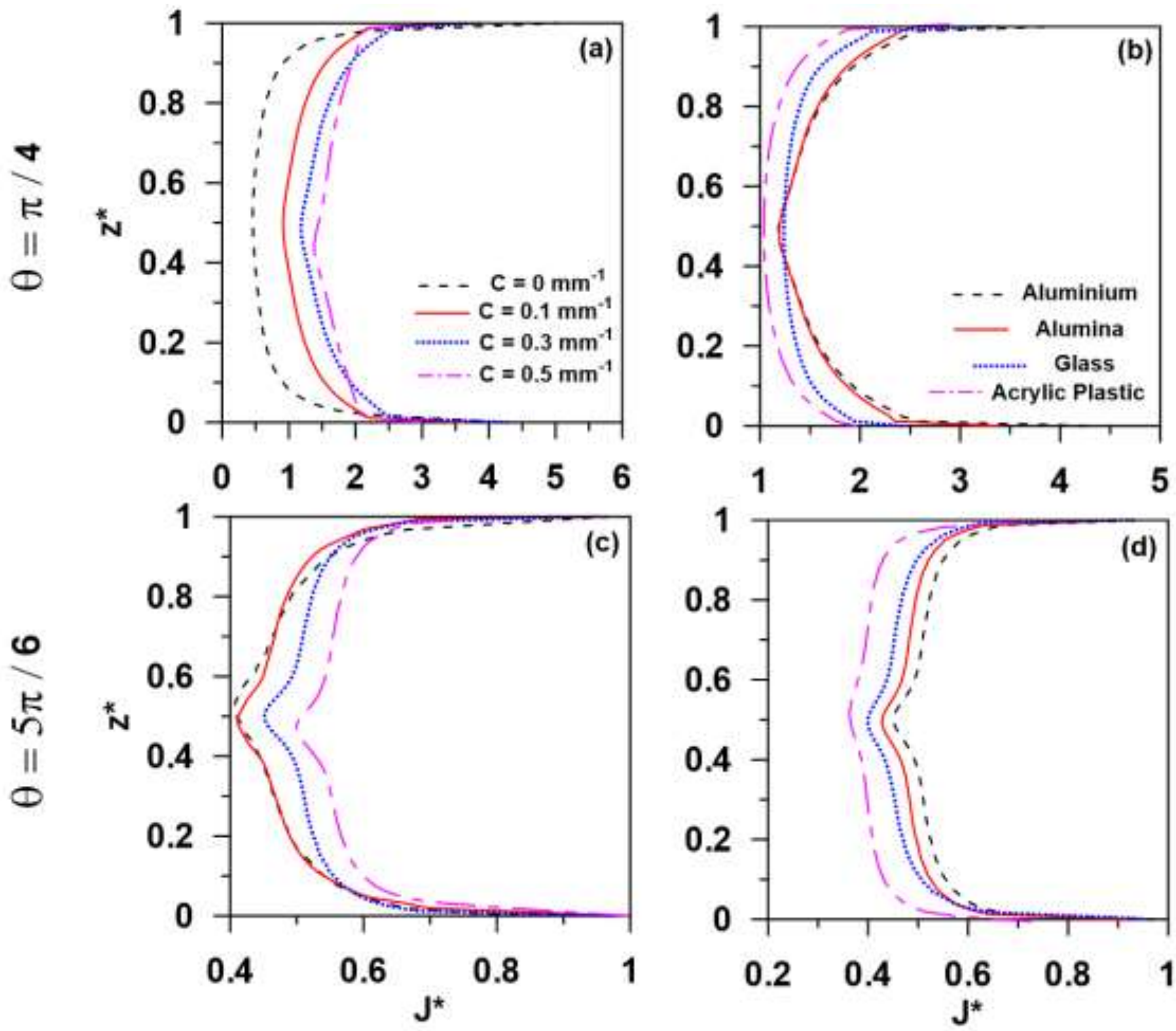

**Fig. 5: Distribution of interfacial mass flux along the capillary surface of liquid bridges under varied curvature and thermal conductivity cases for hydrophilic (a, b) and SH surfaces (c, d) [The legends are same for (a), (c) and (b), (d)].**

In case of curved substrates, we notice that the evaporative mass flux consistently increases with the enhanced convex curvature for both hydrophilic and SH surfaces. Furthermore, for a given solid-liquid geometry, the mass flux is reduced with the decreased thermal conductivity of the solid substrate material. These trends in the evaporative mass flux are consistent with the corresponding evaporation rates and are primarily attributed to the curvature-induced modification of the capillary surface geometry, as discussed earlier. We also notice that, there is a local decrease in evaporative flux around $z^* \approx 0.5$ for all the cases considered in present study. This behaviour can be ascribed to the local reduction in interfacial temperature due to evaporative cooling effects thereby reducing saturation vapor concentration at liquid-vapor interface and consequently weakens the vapor concentration gradient in the surrounding gaseous domain. These effects are more pronounced for SH surfaces with high thermally conductive materials. As a result, the local dip in evaporative flux around $z^* \approx 0.5$ is more prominent for SH surfaces compared to hydrophilic cases.

**3.3 Thermo-fluidic transport behaviour**

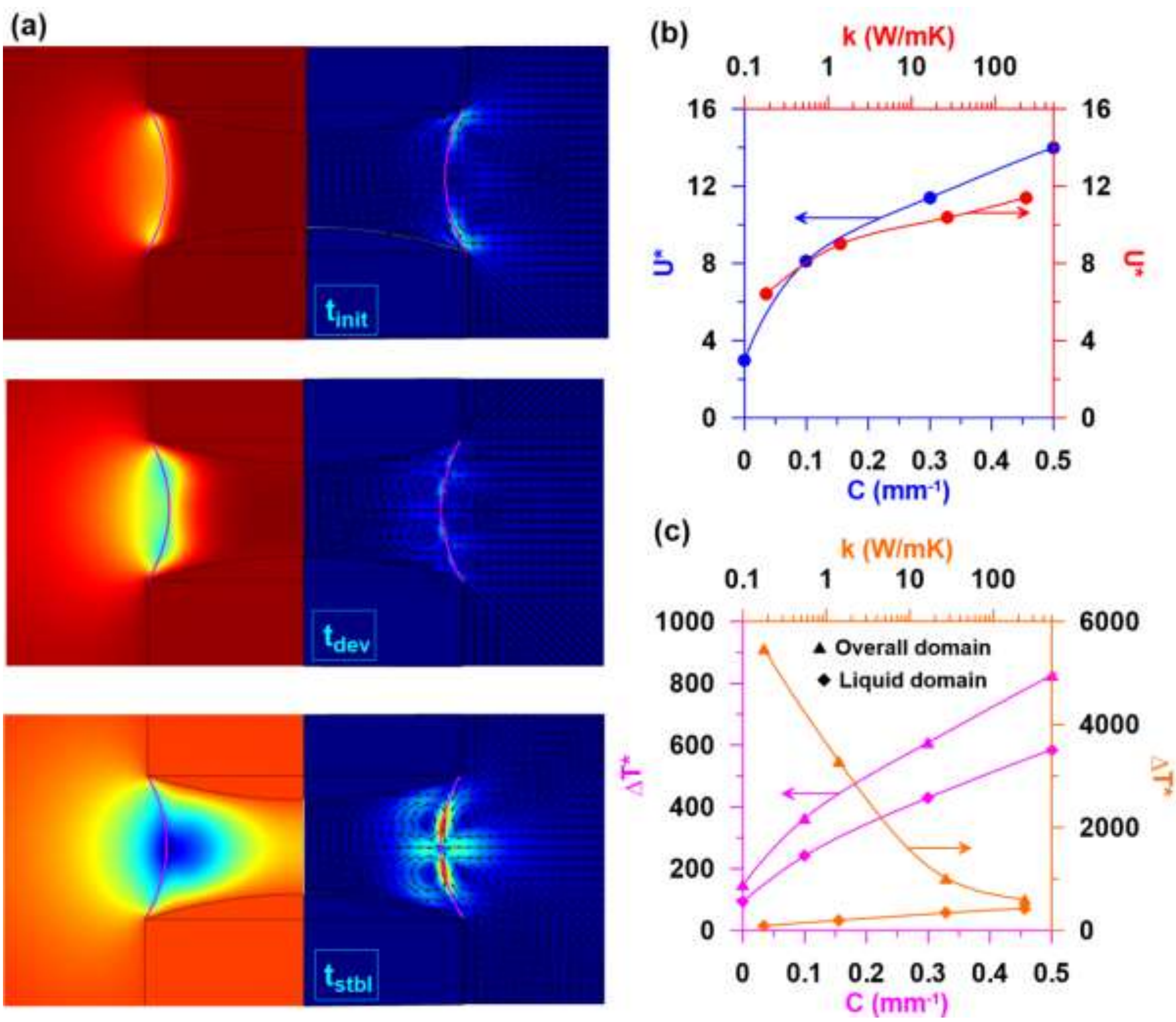

**Fig. 6: (a) Distribution of temperature and velocity field in the liquid and gaseous domain during evaporation of capillary bridges formed between hydrophilic curved substrates; Variation of temperature and velocity scale for different (b) curvature and (c) thermal conductivity of solid substrates.**

The non-uniform evaporative flux results in uneven evaporative cooling effects along capillary surface thereby cause bulk fluid motion in both liquid and gaseous domain due to thermal Marangoni effects or buoyancy driven advection. Fig. 6 (a) illustrates these temperature and velocity field during evaporation for liquid bridge formed between curved hydrophilic surfaces. The subscripts *init, dev, stbl* represent conditions corresponding to initial, developing and stable stages of flow field during evaporation. The corresponding velocity and temperature scales representing peak velocity ($U_p$) and highest temperature drop ($\Delta T_p = T_{amb} - T_{lowest}$) for different curvature and thermal conductivity of solid substrates are shown in fig. 6(b) and 6(c) respectively. These results for SH surfaces are illustrated in fig. 7. In both these plots, the velocity and temperature scales are non-dimensionalized as: $U^* = U_p / (\upsilon_w / R_0)$ and $\Delta T^* = \Delta T_p \big/ (\mu_w \alpha_w / \gamma_T R_0)$; where $\upsilon_w$, $\mu_w$, $\alpha_w$, $\gamma_T$ and $R_0$ represent kinematic and dynamic viscosities, thermal diffusivity of water, temperature derivative of surface tension for air-water interface, and, overall radius of interfacial curvature respectively. Now, before analyzing the thermo-hydrodynamic characteristics, we first examine the underlying physical mechanisms governing the generation of the flow field. In case of thermal imbalance, bulk fluid motion may be generated due to Marangoni or buoyancy effects. To weigh the relative importance of these flow triggering mechanisms, we obtain the Marangoni number and Grashoff number as: $Ma = \gamma_T \Delta T_p R_0 / \mu_w \alpha_w$ and $Gr = \beta g \Delta T_p R_0^3 / \upsilon_w^{\ 2}$. These non-dimensional numbers under varied considered cases are summarized in table 2. Physically, these numbers represent the relative strength of Marangoni effects or buoyance force over viscus effects in generating molecule mobility disturbances. For all considered conditions, we notice that the Marangoni number is significantly higher or in same magnitude order of the critical conditions ($Ma_{CR}$ =83 [46]) for realizing the Marangoni effects. However, the obtained Grashoff numbers are /// order of magnitude smaller than critical state ($Gr_{CR}$=2400 [47, 48]) to trigger buoyance driven motion. Hence, we infer the dominance of Marangoni advection over buoyance effects in setting off bulk fluid motion during evaporation. We also obtain the heat Peclet number $(Pe_h = VR_0 / \alpha_w)$ to probe internal heat transport mechanism as shown in table 2. As observed, the high Peclet numbers for most considered cases represent the strong influence of heat advection over conductive effects in the overall internal heat transfer phenomenon.

**Table 2: Non-dimensional numbers to characterize internal thermo-fluidics**

| Wetting effects | | $\theta = \pi/4$ | | | $\theta = 5\pi/6$ | | |
|---|---|---|---|---|---|---|---|
| Non-dimensional numbers | | *Ma* | *Gr* | $Pe_h$ | *Ma* | *Gr* | $Pe_h$ |
| Effect of | $C_0$ | 93.98 | 0.06 | 20.38 | 112.78 | 0.01 | 25.44 |

| curvature for constant thermal conductivity | $C_1$ | 241.67 | 0.61 | 55.56 | 124.86 | 0.02 | 28.02 |
|---|---|---|---|---|---|---|---|
| | $C_2$ | 429.65 | 1.99 | 78.13 | 169.17 | 0.04 | 36.88 |
| | $C_3$ | 584.05 | 3.43 | 95.79 | 194.57 | 0.05 | 40.54 |
| Effect of thermal conductivity for constant curvature | Aluminum | 429.65 | 1.99 | 78.13 | 169.17 | 0.04 | 36.88 |
| | Alumina | 340.13 | 1.58 | 71.01 | 140.97 | 0.03 | 34.54 |
| | Glass | 196.92 | 0.915 | 61.58 | 68.47 | 0.016 | 22.78 |
| | Acrylic | 89.51 | 0.415 | 43.83 | 64.44 | 0.014 | 18.19 |

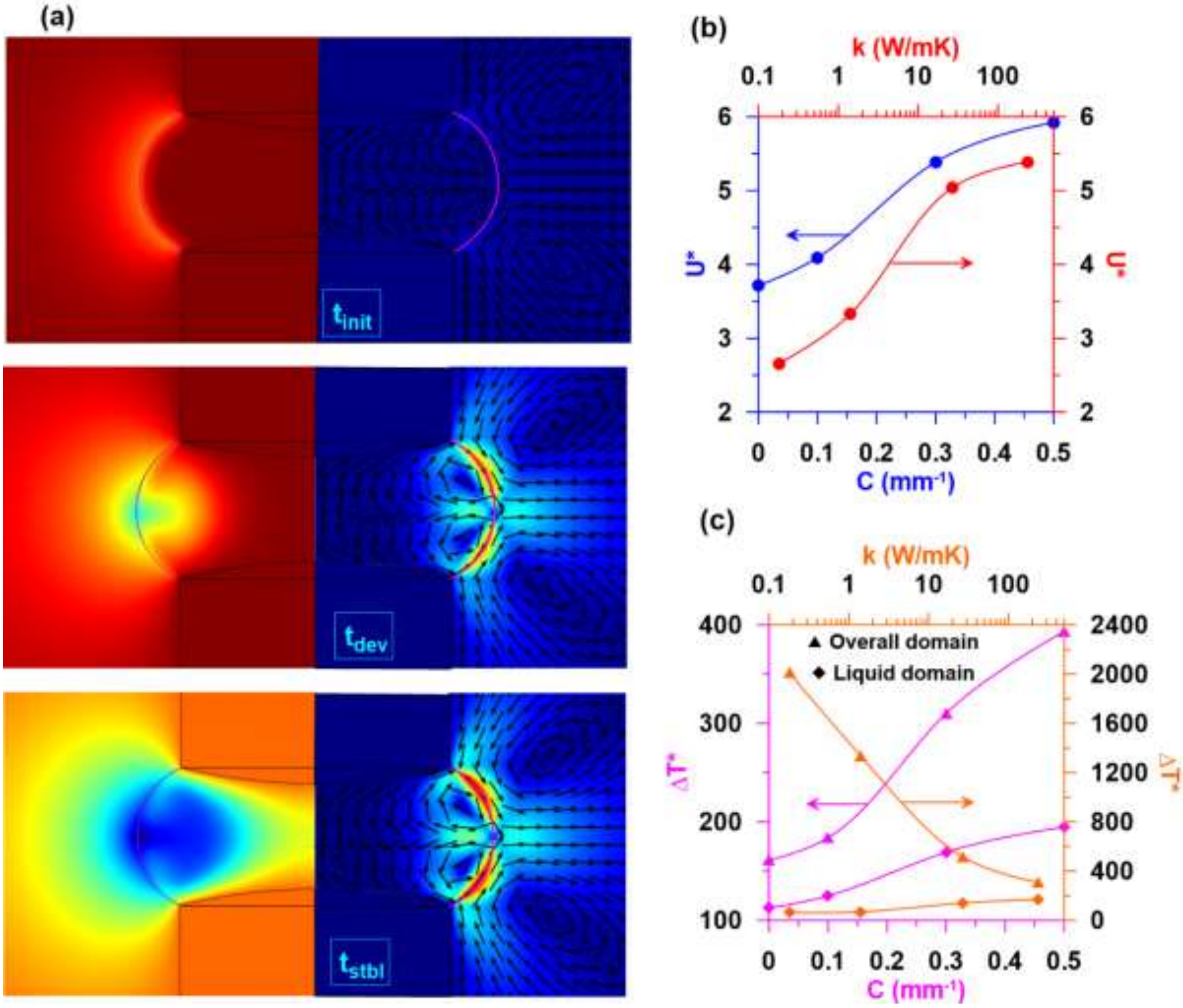

**Fig. 7: (a) Distribution of temperature and velocity field in the liquid and gaseous domain during evaporation of capillary bridges formed between SH curved substrates;**

**Variation of temperature and velocity scale for different (b) curvature and (c) thermal conductivity of solid substrates.**

Next, we look into the transient development pattern of the evaporation induced thermo-flow-field under varied conditions considered in present study. For hydrophilic surfaces, the peak evaporative flux at initial condition locally cools down the fluid edges near three phase contact lines. This condition set off two co-vortices sweeping liquid along liquid-vapor interface from neck region ($z^* \approx 0.5$) towards the contact points as seen in fig. 6(a) at the initial stage. With progressing time, this fluid motion result in internal heat advection thereby resulting in generation of multiple local thermal extrema at the interface. This condition results in formation of multiple local vortices temporarily adjacent the interface during the flow developing stages. Finally, at stable state, the primary cooling spot shifts towards the neck region, resulting in a monotonic temperature variation from the highest value at the three-phase contact line to the lowest value at the neck region. As a result, the stable flow field consist of two counteracting vortices transferring fluid from the contact line towards neck region along interface. In case of SH surfaces (see fig 7(a)), a more uniform mass flux cools the entire interface in a more even manner thereby lowering the liquid domain temperature than the solid. As a result, the counter-rotating vortices become clearly discernible from the onset of evaporation and become increasingly pronounced upon reaching the stable state.

Now, although this evaporation induced thermo-flow-field remain qualitatively similar for altered curvature and thermal conductivity of the solid substrates, the corresponding velocity and temperature scale exhibit significant variations as shown in fig. 6(b, c) and 7(b, c). For instance, we observe that, the velocity scale is consistently increased with enhanced curvature and thermal conductivity of solid substrates. The corresponding temperature drops in both liquid and gaseous domain is intensified consistently with enhanced curvature of solid substrates. However, the overall temperature drop increases in a more rapid manner than that of liquid domain. In other words, a significant fraction of evaporative cooling induced temperature drop remains concentrated in the solid domain for high curvature cases. The trend in these variations of temperature and velocity scales for different curvatures remain qualitatively similar for both hydrophilic and SH surfaces. These findings can be ascribed to the enhanced evaporation rate with curtailed solid-liquid interfacial area for curved geometries. It is can be noticed that (see fig. 3) the solid-liquid interfacial area is shrunk for curved substrates than the flat one thereby limits the heat transfer rate from solid to fluid domain. Whereas, the evaporation rate is increased due to altered liquid-vapor interfacial geometry. Both these factors favor steep thermal gradient in solid domain to compensate for enhanced evaporative cooling effects. As a result, the temperature drop in solid domain is steeply increased than the liquid one for curved geometries of solid substrate. The slower increase in the temperature drop within the liquid domain also explains the corresponding reduction in the rate of increase of the velocity scale as shown in Fig. 6 (b) and 7 (b).

For substrates with altered thermal conductivities, the variation in temperature drops across the solid and liquid domains differs significantly, as shown in Figs. 6(c) and 7(c). Specifically, the overall temperature drop increases sharply with decreasing thermal conductivity of the solid substrate, whereas the temperature drop within the liquid domain decreases slightly. Consequently, although the overall evaporative cooling becomes substantially stronger for substrates with lower thermal conductivity, the corresponding velocity scale is reduced. Apparently, this trend appears counterintuitive, since stronger evaporative cooling is generally expected to enhance the Marangoni effects and thereby upscaling the flow velocity. This seemingly contradictory behavior is primarily attributed to the limited heat conduction capability of thermally insulating substrates. Due to their poor thermal conductivity, a major portion of the evaporative cooling remains confined within the solid domain rather than being transmitted to the liquid. For example, for substrates with very low thermal conductivity (e.g., acrylic plastic, whose thermal conductivity is even lower than that of water), nearly 98% of the overall temperature drop remains concentrated within the solid domain. As a result, despite the substantial increase in the overall evaporative cooling, the temperature drop across the liquid domain is slightly reduced. This condition ultimately weakens the thermal gradient within the liquid thereby suppressing the Marangoni stresses, leading to a reduction in the characteristic flow velocity as seen in fig 6(b) and 7(b).

### 3.4 Adhesion Dynamics:

The evolving liquid-vapor interface of liquid bridges during evaporation can significantly influence the capillary force interaction between the two solid cylinders. Physically, the capillary force comprises two components: the pressure force ( $F_{\Delta P}$ ) arising from the pressure difference across the liquid-vapor interface and the surface tension force ( $F_{ST}$ ) acting along the contact line (or gorge perimeter) of the liquid bridge. Mathematically, the total capillary force can be expressed as [49]:

$$F_{cap} = F_{\Delta P} + F_{ST} \tag{24}$$

$$F_{\Delta P} = -\pi r_g^{\,2} \Delta p \tag{25}$$

$$F_{ST} = 2\pi r_g \gamma \tag{26}$$

Here, $r_g$ is the gorge radius and $\Delta p$ is the pressure difference across the liquid-vapor interface. These classical expressions for pressure and surface tension component of capillary force are derived under the fundamental assumption that the liquid-vapor interface remains in mechanical equilibrium. Since the interface continuously evolves during evaporation in the present study, it is worthwhile to first examine the validity of these expressions under dynamic conditions. In present study, the evaporation induced flow velocity for all cases remain below the order $O$ (10) mm/s. The corresponding dynamic pressure component becomes $p_d \approx 0.5\rho U^2 \sim O(10^{-2})$ Pa which is much smaller than the capillary pressure component ($p_c \sim O(10^2)$) of meso-scale capillary bridge [28, 29]. In addition, to assess the influence of viscous effects relative to surface tension in modifying the liquid-vapor interface through the interfacial stress balance, we evaluate the capillary number as $Ca = \mu_w U_p / \gamma$. In present study, we observed that the capillary number for all test cases remain in the order $Ca \sim O(10^{-4})$. Thus, we note that surface tension strongly dominates over viscous effects at liquid-vapor interface. Also, in case of water droplet evaporation the process follows quasistatic equilibrium behaviour [50]. Therefore, we infer that interface evolves in a quasi-static manner and remains close to mechanical equilibrium throughout the evaporation process. Consequently, the classical expressions for the capillary force calculation remain valid throughout the considered stages of evaporation. In present study, we directly employ the numerically obtained average pressure with in liquid domain and surface tension at gorge perimeter in eqn. (25) and (26) to obtain the capillary force components at different stages of evaporation.

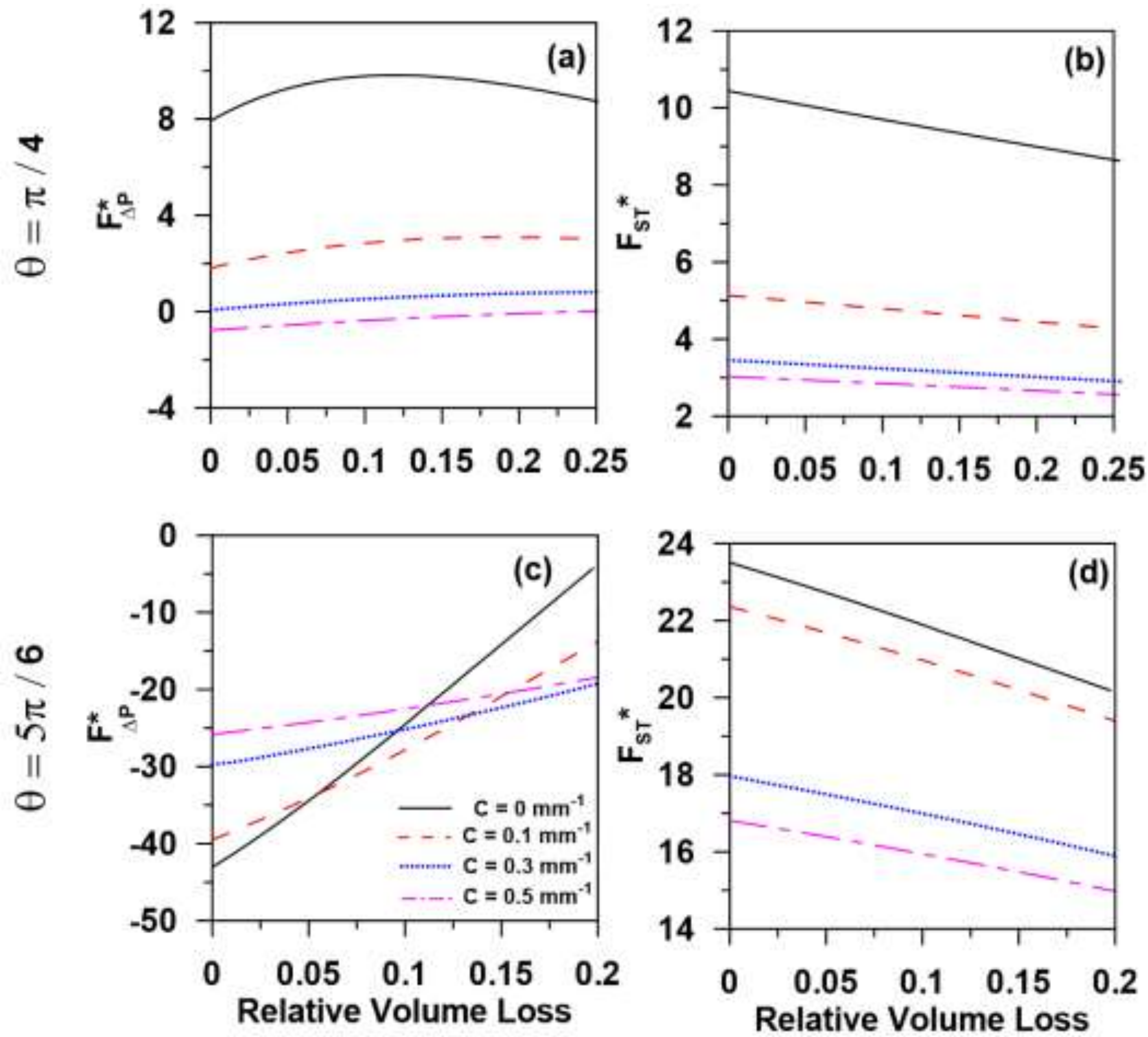

**Fig. 8: Dynamic evolution of pressure and surface tension component of capillary forces due to relative volume loss of evaporating capillary bridges for hydrophilic (a, b) and SH surfaces (c, d) [The legends are same for all sub-figures].**

The dynamic evolution of pressure ($F_{\Delta P}$) and surface tension ($F_{ST}$) component of capillary forces due to relative volume loss of capillary bridges during evaporation are illustrated in fig. 8. The corresponding total capillary force for varied curvature and thermal conductivity of solid substrates are shown in fig. 9. Here, the force components are non-dimensionalized as: $F_{ST}* = F_{ST}/\gamma R_0$, and $F_{\Delta P}* = F_{\Delta P}/\gamma R_0$. In these plots an attraction force is considered as positive while the repulsive one is assumed as negative. We notice that for hydrophilic surface the capillary pressure increases to a maximum value and then decreases with progression the progressing volume loss under varied substrate curvature during evaporation. Also, the overall magnitude of capillary pressure is consistently reduced for curved geometry cases than the flat substrate. These trends can be attributed to the transient evolution of the liquid-vapor interfacial geometry during evaporation. At the initial stages, the external radius of curvature decreases more rapidly than the gorge radius. Consequently, the capillary suction pressure increases, although its rate of increase gradually diminishes [29]. As evaporation proceeds, the rates of change of the external and gorge radii become nearly identical, resulting in an instability point: $d(\Delta p)/dV = 0$ i.e., the suction pressure remains nearly constant although the volume changes. Beyond this stage, the gorge radius decreases at a higher rate than the external radius of curvature, leading to a progressive reduction in the capillary suction pressure with continued liquid volume loss [29]. Also, the overall reduction in capillary curvature justifies the smaller magnitude of internal capillary pressure for curved substrates as compared to the flat one.

In case of SH surfaces, the large contact angle at solid-liquid interface results in internal gauge pressure thereby resulting in repulsive capillary pressure force between the solid substrates for all cases as shown in fig. 8(c). This initial repulsive force is significantly higher for flat substrates and the pressure intensity is reduced in a prominent manner with enhanced solid substrate curvature due to reduced liquid-vapor interfacial curvature similar to hydrophilic cases. However, with progressive volume loss the internal capillary gauge pressure mitigates at a faster rate for low curvature cases and the rate is diminished with enhanced curvature of the solid substrates. This behavior can be attributed to the curvature-induced modification of the capillary bridge geometry for SH surfaces. With increasing substrate curvature, a larger fraction of the liquid bridge protrudes into the gap between the solid cylinders, causing the liquid-vapor interface to shift farther away from the vertical plane passing through the three-phase contact line (see fig. 3). Consequently, a greater amount of liquid must evaporate to produce an equivalent change in internal capillary pressure. This delays the reduction in the capillary gauge pressure, resulting in a slower rate of pressure decay during evaporation over curved substrates.

We further observe that the surface tension component of the capillary force remains positive and decreases monotonically throughout the evaporation process for both hydrophilic and SH surfaces. This monotonic decrease is primarily attributed to the progressive reduction in the gorge radius during evaporation. The surface tension force remains positive because the liquid-vapor interface continuously exerts a tensile force along the contact line, thereby attracting the two solid substrates toward each other. However, the total capillary force is determined by the combined contributions of the pressure and surface tension components, as shown in Fig. 9. For hydrophilic surfaces, the pressure component is also attractive (suction), and therefore both components act synergistically, resulting in an overall attractive capillary force. In contrast, for SH surfaces, the positive internal gauge pressure generates a repulsive pressure force that exceeds the attractive surface tension component. Consequently, the resultant capillary force becomes repulsive. For substrates with lower thermal conductivity, the overall capillary force magnitude is marginally enhanced for both hydrophilic and SH surfaces, as shown in Fig. 9 (c, d). This behavior is primarily attributed to the increase in surface tension resulting from the lower interfacial temperature caused by the enhanced evaporative cooling. As the thermal conductivity of the substrate decreases, the intensified evaporative cooling reduces the liquid-vapor interfacial temperature, thereby increasing the surface tension. Consequently, the both the surface tension and pressure component of the capillary force is consistently enhanced, leading to a slight increase in the overall capillary force magnitude with decreasing substrate thermal conductivity.

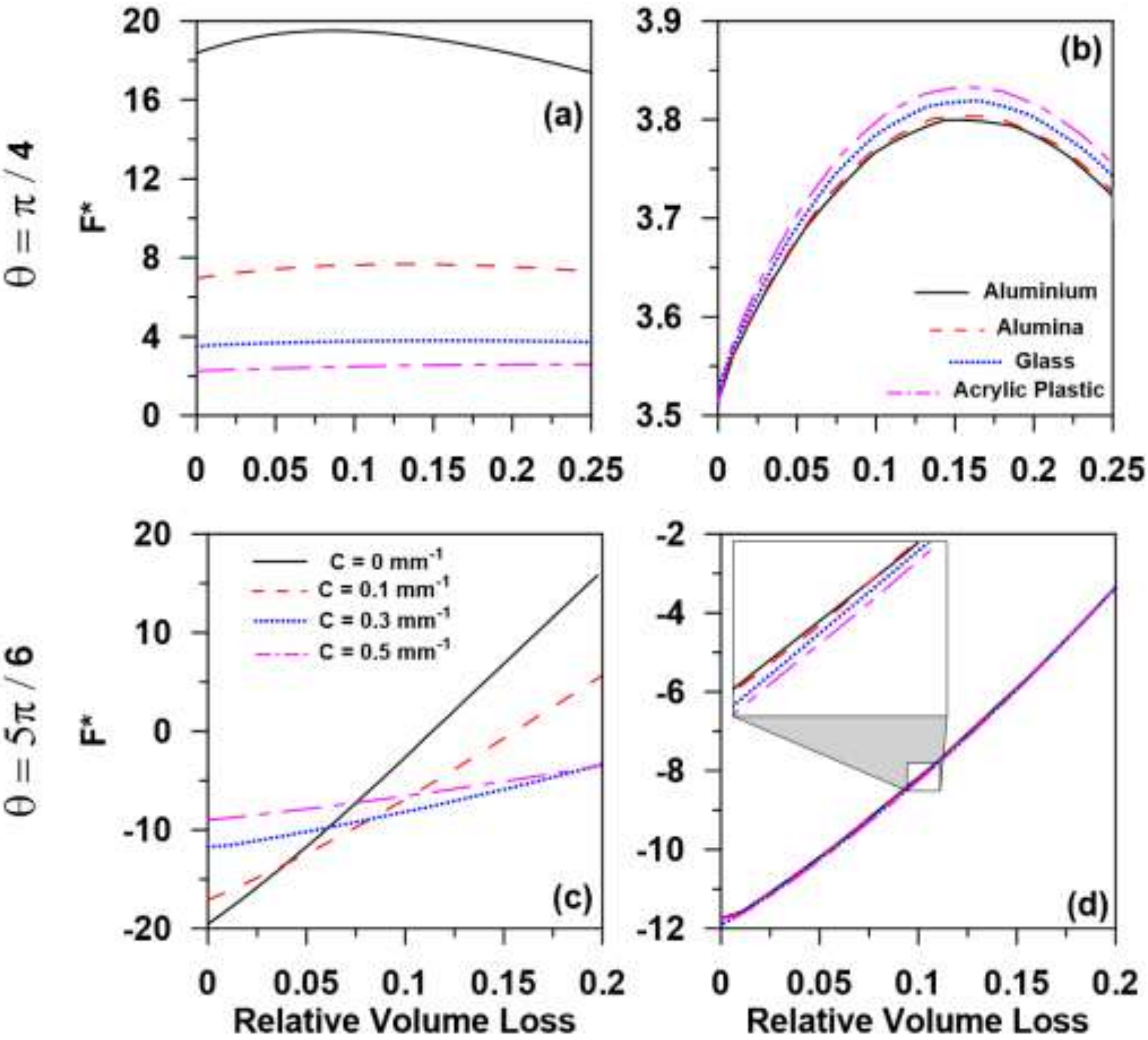

**Fig. 9: Dynamic evolution of total capillary force due to relative volume loss of evaporating capillary bridges for varied curvature and thermal conductivity of solid substrates [The legends are same for (a, c) and (b, d)].**

## 4. Conclusions

To conclude, we probe the thermo-hydrodynamics, species transport, and adhesion dynamics of evaporating capillary bridges under the combined effects of substrate curvature, wettability, and thermal conductivity. Equilibrium capillary bridge profiles are first obtained using the level-set method and subsequently employed as the initial configuration for evaporation simulations based on a transient, fully coupled Arbitrary Lagrangian-Eulerian (ALE) framework. The numerical predictions show good agreement with available experimental and theoretical studies, demonstrating the accuracy of the proposed model. The key findings of the present investigation are summarized as:

- For both hydrophilic and superhydrophobic (SH) surfaces, increasing the convexity of adjacent substrates significantly enlarges the liquid-vapor interfacial area and its radius of curvature, thereby enhancing the evaporative mass loss owing to the increased mass transfer area and steeper vapor concentration gradient at the interface. In contrast, substrates with lower thermal conductivity suppress the evaporation rate by limiting heat conduction to the liquid bridge, resulting in lower interfacial temperatures and a consequent reduction in the local saturation vapor concentration at the interface.
- The presence of solid boundary dictates diverging evaporative flux near the three-phase contact line. This non-uniformity in mass flux results in uneven evaporative cooling effects thereby set off bulk internal motion. The non-dimensional scaling analysis showed that the Marangoni effects strongly dominate over buoyancy effects in the overall flow features. This Marangoni-induced interfacial flow sweeps liquids from the contact line towards the gorge region thereby resulting in formation of two counter-acting vortices at the stable state. Moreover, this bulk internal flow causes heat advection thereby affecting the overall thermo-fluid-species transport phenomena during the evaporation process.
- The evaporation induced velocity scales are significantly higher for curved substrates due to enhanced evaporative cooling caused by higher evaporation rate. However, the velocity scales are diminished for liquid bridges formed between low thermally conductive surfaces despite substantially higher evaporative cooling effects. This behavior is attributed to the low thermal conductance of the solid substrate, which confines nearly 98% of the total evaporation-induced temperature drop within the substrate. Consequently, the thermal gradient within the liquid domain is substantially weakened, thereby suppressing the Marangoni stresses and the associated flow velocities.

- The net capillary force is predominantly positive (suction) for hydrophilic surfaces, whereas it remains predominantly negative (repulsion) for superhydrophobic (SH) surfaces during most stages of evaporation owing to the altered wetting characteristics. Induction of substrate convexity, reduces the overall magnitude of capillary force due to diminished capillary curvature. Also, a reduction in the thermal conductivity of the solid substrate slightly enhances the capillary force due to the marginal increase in surface tension resulting from the lower interfacial temperature.

**Data availability**: All data pertaining to the research are in the article.

**Conflicts of interest**: The authors declare having no conflicts of interests.

**Acknowledgements**: AP and PD thank MNNIT Allahabad and IIT Kharagpur for partially funding the research via internal grants.

## Refferences: